# Audio Compression using Graph-based Transform

1st Majid Farzaneh
*Faculty of Media Technology and Engineering*
*Iran Broadcasting University*
Tehran, Iran
Majid.Farzaneh91@gmail.com

2nd Mohammad Asgari and 3rd Rahil Mahdian Toroghi
*Faculty of Media Technology and Engineering*
*Iran Broadcasting University*
Tehran, Iran
{asgarimohammad, mahdian.t.r}@gmail.com

*Abstract*— Graph-based Transform is one of the recent transform coding methods which has been used in some state-of-arts for data decorrelation. In this paper, we propose a Graph-based Transform (GT) for audio compression. In this method we introduce a proper graph structure for audio. Then audio frames are projected onto an orthogonal matrix consisting of eigenvectors of the introduced graph matrix, leading to sparse coefficients. The results show that the proposed method can de-correlate audio signal better than other transform methods like Discrete Cosine Transform (DCT) and Walsh-Hadamard Transform (WHT).

*Keywords*— *Graph-based Transformation (GT), Audio compression, Audio coding, Discrete Cosine Transform (DCT), Decorrelation.*

## I. INTRODUCTION

Audio compression is an ongoing project in the audio engineering community which allows the efficient storage and transmission of audio data. Lossy audio compression is used in a wide range of applications, such as digital audio streams on the Internet, and terrestrial radio broadcasts.

The compression performance is typically measured by the complexity level of the algorithm, the quality of the compressed audio, as well as the amount of which the audio data is compressed [1].

There is a plethora of lossless and lossy compression techniques within this context, most of which are standardized and specifically coined in the commercial digital audio community. MPEG/audio compression is one part of a larger MPEG standard which integrates audio, video and synchronization of them to an aggregate bit rate of 1.5 Mbits/sec.

Lossy compression methods convert the temporal sampled audio waveform into a transform domain (normally the spectral domain) in order to de-correlate the frequency components and allocate bits to them according to their audibility level. MPEG/audio, as one of the most prevalently used compression techniques, incorporates the Modified Discrete Cosine Transform (MDCT) [2] for a lossless sub-band transformation, and bit allocation.

Recently, the Graph-based Transform (GT) has grabbed the attention of researchers in image and video coding context [3-5], in which the Discrete Cosine Transform (DCT) [6] is replaced by an edge-adaptive transform for an efficient depth map coding. Due to the piecewise smooth nature of the video signals, for coding applications, adaptive selecting the transform or sparsifying basis and then signaling the selected transform to the decoder often achieves better performance than using an *a priori* known sparsifying basis (e.g. DCT) at the decoder. It has been shown that for piecewise smooth signals, where sharp edges exist between smooth regions, edge-adaptive transforms introduce sparser representation, hence better compression rates could be achieved.

Alongside the before-mentioned vision, here we introduce a Graph-based algorithm which is developed for exploring the correlation of the signals with first-order Markov structure (e.g. audio signals).

The novelty of this paper explicitly consists of leveraging the GT for audio compression and introducing a proper graph structure for audio signals.

In [5], the GT was used to compress human motion capture data and showed that GT performed better than DCT compression. Adaptive graph-based transforms (EA-GBT) are used to video coding. It is also concluded in [4] that the DCT acts poorly in the face of some data, such as human motion in video, and is not able to de-correlate properly.

The rest of this paper is organized as follows: Section II introduces GT and its background. Section III introduces the appropriate graph structure and how to compress audio using GT. In Section IV, the test results of the proposed method are presented on two speech and music data sets, and finally, a conclusion is made in Section V.

## II. GRAPH-BASED TRANSFORMATION

Given a block of an audio signal with a frame size of *N* samples, we can create a graph *G={V,E,s}* where *V* and *E* are the vertices and edges of the graph, and $s \epsilon \mathbb{R}^{N\times 1}$ is an audio signal for which the graph matrix is defined as $K \epsilon \mathbb{R}^{N\times N}$. For this graph, the adjacency matrix W elements are obtained as

$$W_{ij} = \begin{cases} w_{ij}. & if\ (i.j) \in E \\ 0. & otherwise \end{cases} \quad (1)$$

where $w_{ij}$ is the weight of the edge between *i and j* in the graph. The degree matrix $D \epsilon \mathbb{R}^{N\times N}$ is a diagonal matrix, for which the elements are defined as follows,

$$D_{ij} = \begin{cases} \sum w_{ij}. & if\ i = j \\ 0. & otherwise \end{cases} \quad (2)$$

Then, the Graph-Laplacian Matrix *K* would be defined as,

$$K = D - W \quad (3)$$

where the operator *K* is also known as *Kirchoff* operator, as a tribute to Gustav Kirchoff for this studies and achievements on electrical networks. Kirchoff referred to the (weighted) adjacency matrix *W* as the *conductance* matrix.

The matrix *K* would be a real symmetric one, and based upon the *spectral theory*, the eigenvalue decomposition (EVD) of this matrix would lead to a set of real non-negative eigenvalues, denoted by $\Lambda = \{\lambda_1. ... . \lambda_N\}$, and a set of corresponding independent (hence, orthogonal) eigenvectors denoted by $V = \{v_1. ... . v_N\}$, derived as,

$$K = V\Lambda V^T \qquad (4)$$

We can then use these orthogonal eigenvectors to de-correlate the signal defined on the graph, i.e.,

$$c = V^T s \qquad (5)$$

where $c \epsilon \mathbb{R}^{N \times 1}$ is the approximate sparse transform coefficient matrix. For more information, please see [7] and [8].

### III. GT-BASED AUDIO COMPRESSION

To apply the GT on the input audio signal, we have introduced two graph structures, as in Figure1 and Figure2. Considering the fact that near samples of an audio signals would be highly correlated, we assume any sample in a frame as one node in the graph that interconnects with its neighborhood samples through the edges. For simplicity, the explained structure is revealed in Figure1 for an example of a frame with 8 samples length.

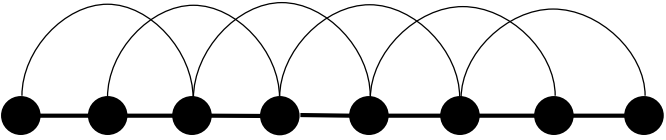

**Figure 1. Graph structure I**

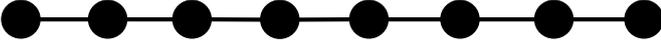

**Figure 2. Graph structure II**

The adjacency matrices could be defined as follows:

$$W_1 = \begin{bmatrix} 0 & 1 & 0.1 & 0 & 0 & 0 & 0 & 0 \\ 1 & 0 & 1 & 0.1 & 0 & 0 & 0 & 0 \\ 0.1 & 1 & 0 & 1 & 0.1 & 0 & 0 & 0 \\ 0 & 0.1 & 1 & 0 & 1 & 0.1 & 0 & 0 \\ 0 & 0 & 0.1 & 1 & 0 & 1 & 0.1 & 0 \\ 0 & 0 & 0 & 0.1 & 1 & 0 & 1 & 0.1 \\ 0 & 0 & 0 & 0 & 0.1 & 1 & 0 & 1 \\ 0 & 0 & 0 & 0 & 0 & 0.1 & 1 & 0 \end{bmatrix}$$

$$W_2 = \begin{bmatrix} 0 & 1 & 0 & 0 & 0 & 0 & 0 & 0 \\ 1 & 0 & 1 & 0 & 0 & 0 & 0 & 0 \\ 0 & 1 & 0 & 1 & 0 & 0 & 0 & 0 \\ 0 & 0 & 1 & 0 & 1 & 0 & 0 & 0 \\ 0 & 0 & 0 & 1 & 0 & 1 & 0 & 0 \\ 0 & 0 & 0 & 0 & 1 & 0 & 1 & 0 \\ 0 & 0 & 0 & 0 & 0 & 1 & 0 & 1 \\ 0 & 0 & 0 & 0 & 0 & 0 & 1 & 0 \end{bmatrix}$$

We have used the first structure for GT-I and the second one for GT-II method. As shown in Figure1, weights of edges between a sample and its first neighbor is larger than weight of second neighbor. Figure3 shows an example result of using GT-I on a speech frame with 512 samples.

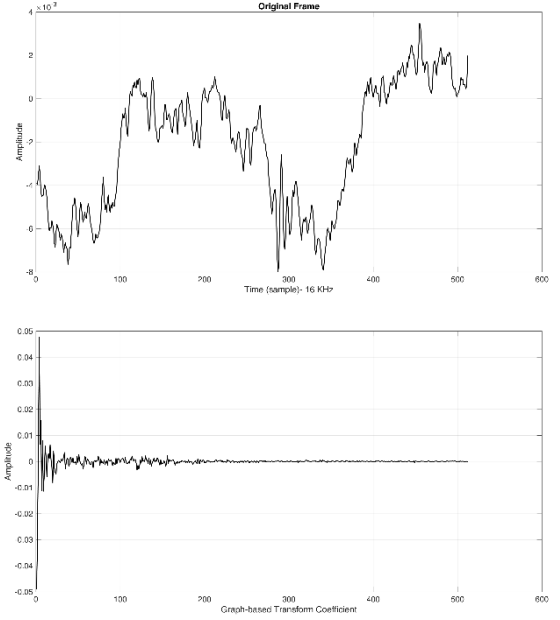

**Figure 3. An example of GT**

### IV. EXPERMINTAL RESULTS AND DISCUSSIONS

We have tested the GT-I and GT-II methods on weather speech dataset of Carnegie Mellon University's speech group[1] and also 10 music in different genres. We compared results with DCT-I [6], DCT-II [9], DCT-III [10], DCT-IV [11] and Fast Walsh-Hadamard Transform (FWHT) [14] based on Peak Signal to Noise Ratio (PSNR) value, Energy Retained Percentage (ERP), Short-Time Objective Intelligibility (STOI) [12] and Perceptual Evaluation of Speech Quality (PESQ) [13]. We have considered several frame size and several compression ratio (CR). The results shown in Table1 to Table6.

In Table1, the results of transforming with different conditions are displayed on 40 speech signals with bitrate of 256. In the GT-I, the weight of the edge between each sample and its first neighbor is equal to 1 and for the edge between each sample with its second neighbor is 0.1. There are no edges between the other samples. For GT-II, each sample has its edge with its first neighbor and the weight value is 1. As you can see, DCT-II's results with GT-II are exactly equal. Because the GT-I's matrix is very similar to the DCT-II basis matrix. This result is also seen in other evaluations in Tables 2, 3 and 4. In some cases, the GT-I has shown better results, and it shows that it is capable of better results from DCT types. In all tables, FWHT is not a good way to compress audio.

In Table2, the average retained energy percentage is displayed for 40 speech signals. The results show that GT have the highest energy after compression compared to other methods.

Table3 shows the average results of the PSNR value, by executing different methods on 10 music signals from different genres. In this table, the superiority of the GT-I is well seen. It should be noted that since samples in the musical signals are more similar to their neighbors, we assume that the adjacency weights for each sample with the first neighbor, is 1 and for second neighbor is 0.3. By comparing Table 1 and 3, GT-I method is better for musical signals than for speech. Table 4 also shows that the energy retained in GT-I is more than other methods after compression.

---

[1] Available at http://festvox.org/dbs/dbs_weather.html

**Table 1. PSNR (dB) (mean value for 40 speech signals)**

| Frame Size | CR | GT-I | GT-II | DCT-I | DCT-II | DCT-III | DCT-IV | FWHT |
|---|---|---|---|---|---|---|---|---|
| 16 | 2:1 | 40.6580 | 40.7288 | 38.7618 | 40.7288 | 38.4002 | 35.6453 | 35.7558 |
| | 4:1 | 32.3845 | 32.3856 | 32.0461 | 32.3856 | 31.0529 | 30.1466 | 30.4149 |
| | 8:1 | 27.6285 | 27.6292 | 27.7268 | 27.6292 | 27.0142 | 26.6035 | 26.7864 |
| | 16:1 | 23.8039 | 23.8039 | 23.8638 | 23.8039 | 23.4458 | 23.2941 | 23.8039 |
| 64 | 2:1 | 42.1652 | 42.1848 | 41.2856 | 42.1848 | 41.1077 | 39.5834 | 35.7546 |
| | 4:1 | 32.9671 | 32.9664 | 32.8393 | 32.9664 | 32.4807 | 32.1423 | 30.4137 |
| | 8:1 | 28.7965 | 28.7964 | 28.7865 | 28.7964 | 28.4502 | 28.3031 | 26.7854 |
| | 16:1 | 24.1722 | 24.1723 | 24.2197 | 24.1723 | 24.2825 | 24.2402 | 23.8027 |
| 256 | 2:1 | 42.5763 | 42.5657 | 62.2566 | 42.5657 | 42.1982 | 41.6492 | 35.7515 |
| | 4:1 | 33.0741 | 33.0738 | 33.0393 | 33.0738 | 32.9375 | 32.8386 | 30.4107 |
| | 8:1 | 29.0110 | 29.0135 | 29.0111 | 29.0111 | 28.9296 | 28.8882 | 26.7821 |
| | 16:1 | 24.5073 | 24.5073 | 24.5155 | 24.5073 | 24.4859 | 24.4746 | 23.7996 |
| 512 | 2:1 | 42.6339 | 42.6274 | 42.4670 | 42.6274 | 42.4363 | 42.1476 | 35.7495 |
| | 4:1 | 33.0876 | 33.0875 | 33.0718 | 33.0875 | 33.0216 | 32.9707 | 30.4086 |
| | 8:1 | 29.0606 | 29.0607 | 29.0624 | 29.0607 | 29.0215 | 29.0002 | 26.7801 |
| | 16:1 | 24.5422 | 24.5422 | 24.5450 | 24.5422 | 24.5220 | 24.5164 | 23.7976 |

**Table 2. Energy Retained (%) (mean value for 40 speech signals)**

| Frame Size | CR | GT-I | GT-II | DCT-I | DCT-II | DCT-III | DCT-IV | FWHT |
|---|---|---|---|---|---|---|---|---|
| 16 | 2:1 | 99.06 | 99.08 | 98.63 | 99.08 | 98.51 | 97.24 | 97.29 |
| | 4:1 | 94.08 | 94.09 | 93.62 | 94.09 | 92.03 | 90.21 | 90.79 |
| | 8:1 | 82.52 | 82.53 | 82.92 | 82.53 | 79.91 | 77.93 | 78.82 |
| | 16:1 | 57.94 | 57.94 | 58.52 | 57.94 | 54.38 | 52.76 | 57.94 |
| 64 | 2:1 | 99.32 | 99.32 | 99.20 | 99.32 | 99.17 | 98.86 | 97.29 |
| | 4:1 | 94.80 | 94.80 | 94.65 | 94.80 | 94.21 | 93.76 | 90.79 |
| | 8:1 | 86.62 | 86.62 | 86.59 | 86.62 | 85.53 | 85.03 | 78.82 |
| | 16:1 | 61.31 | 61.31 | 61.73 | 61.31 | 62.29 | 61.93 | 57.94 |
| 256 | 2:1 | 99.38 | 99.38 | 99.34 | 99.38 | 99.33 | 99.26 | 97.29 |
| | 4:1 | 94.92 | 94.92 | 94.88 | 94.92 | 94.77 | 94.65 | 90.79 |
| | 8:1 | 87.26 | 87.27 | 87.27 | 87.27 | 87.03 | 86.91 | 87.82 |
| | 16:1 | 64.16 | 64.16 | 64.23 | 64.16 | 63.99 | 63.90 | 57.94 |
| 512 | 2:1 | 99.39 | 99.38 | 99.37 | 99.38 | 99.36 | 99.33 | 97.29 |
| | 4:1 | 94.94 | 94.94 | 94.92 | 94.94 | 94.86 | 94.81 | 90.79 |
| | 8:1 | 87.41 | 87.41 | 87.42 | 87.41 | 87.30 | 87.24 | 78.82 |
| | 16:1 | 64.45 | 64.45 | 64.48 | 64.45 | 64.30 | 64.25 | 57.94 |

**Table 3. PSNR (dB) (mean value for 10 music signals)**

| Frame Size | CR | GT-I | GT-II | DCT-I | DCT-II | DCT-III | DCT-IV | FWHT |
|---|---|---|---|---|---|---|---|---|
| 16 | 2:1 | 60.6079 | 57.2552 | 41.5404 | 57.2552 | 40.5253 | 36.5492 | 44.7925 |
| | 4:1 | 46.3757 | 46.0887 | 38.4973 | 46.0887 | 34.3244 | 32.4676 | 38.3509 |
| | 8:1 | 36.5755 | 36.5640 | 34.6090 | 36.5640 | 29.8139 | 28.9936 | 33.1292 |
| | 16:1 | 28.7137 | 28.7137 | 28.5374 | 28.7137 | 25.7869 | 25.3597 | 28.7137 |
| 64 | 2:1 | 64.9169 | 61.0060 | 46.9724 | 61.0060 | 46.0602 | 42.0435 | 44.7925 |
| | 4:1 | 50.1444 | 49.7648 | 43.1374 | 49.7648 | 39.3772 | 37.6031 | 38.3508 |
| | 8:1 | 40.2980 | 40.2739 | 38.2130 | 40.2739 | 33.9903 | 33.2734 | 33.1291 |
| | 16:1 | 32.1913 | 32.1960 | 31.9826 | 32.1960 | 29.5351 | 29.2640 | 28.7136 |
| 256 | 2:1 | 66.8039 | 63.5491 | 52.0204 | 63.5491 | 51.1751 | 47.3975 | 44.7924 |
| | 4:1 | 52.3933 | 52.0417 | 46.8114 | 52.0417 | 43.7514 | 42.2670 | 38.3506 |
| | 8:1 | 42.0460 | 42.0311 | 40.6798 | 42.0311 | 37.5318 | 36.9790 | 33.1288 |
| | 16:1 | 33.4661 | 33.4657 | 33.3575 | 33.4657 | 31.9857 | 31.8324 | 28.7132 |
| 512 | 2:1 | 67.3550 | 64.5573 | 54.3650 | 64.5573 | 35.5899 | 50.0341 | 44.7923 |
| | 4:1 | 53.2051 | 52.8955 | 48.2999 | 52.8955 | 45.6170 | 44.3060 | 38.3505 |
| | 8:1 | 42.4180 | 42.4084 | 41.4754 | 42.4084 | 38.9511 | 38.4880 | 33.1286 |
| | 16:1 | 33.6115 | 33.6113 | 33.5524 | 33.6113 | 32.7053 | 32.6023 | 28.7129 |

By comparing the results in different frame sizes, the best results are achieved with 512 samples per frame. This frame size also has a shorter execution time than others. Because it requires less conversion and inversion operation.

Table5 shows the mean values of the STOI for the 40 speech signals. Also Table6 shows the mean values of PESQ for these signals. Results show again the GT method has better performance against other methods.

**Table 4. Energy Retained (%) (mean value for 10 music signals)**

| Frame Size | CR | GT-I | GT-II | DCT-I | DCT-II | DCT-III | DCT-IV | FWHT |
|---|---|---|---|---|---|---|---|---|
| 16 | 2:1 | 99.50 | 99.55 | 99.07 | 99.55 | 98.90 | 97.66 | 99.04 |
| | 4:1 | 98.21 | 98.21 | 97.51 | 98.21 | 95.58 | 93.76 | 96.80 |
| | 8:1 | 94.11 | 94.11 | 93.66 | 94.11 | 88.30 | 86.32 | 92.12 |
| | 16:1 | 82.67 | 82.67 | 82.45 | 82.67 | 72.33 | 69.82 | 82.67 |
| 64 | 2:1 | 99.60 | 99.63 | 66.52 | 99.63 | 99.48 | 99.16 | 99.04 |
| | 4:1 | 98.47 | 98.47 | 98.29 | 98.47 | 97.80 | 97.31 | 96.80 |
| | 8:1 | 95.40 | 95.40 | 95.23 | 95.40 | 93.75 | 93.21 | 92.12 |
| | 16:1 | 88.64 | 88.64 | 88.61 | 88.64 | 85.71 | 85.15 | 82.67 |
| 256 | 2:1 | 99.65 | 99.66 | 99.63 | 99.66 | 99.62 | 99.54 | 99.04 |
| | 4:1 | 98.52 | 98.52 | 98.48 | 98.52 | 98.35 | 98.23 | 96.80 |
| | 8:1 | 95.68 | 95.68 | 95.64 | 95.68 | 95.26 | 95.13 | 92.12 |
| | 16:1 | 89.84 | 98.84 | 89.84 | 98.84 | 98.11 | 88.96 | 82.67 |
| 512 | 2:1 | 99.66 | 99.67 | 99.65 | 99.67 | 99.65 | 99.61 | 99.04 |
| | 4:1 | 98.53 | 98.53 | 98.51 | 98.53 | 98.44 | 98.38 | 96.80 |
| | 8:1 | 95.70 | 95.70 | 95.68 | 95.70 | 95.50 | 95.43 | 92.12 |
| | 16:1 | 89.95 | 89.95 | 89.95 | 89.95 | 89.64 | 89.56 | 92.67 |

**Table 5. STOI (mean value for 40 speech signals)**

| Frame Size | CR | GT-I | GT-II | DCT-I | DCT-II | DCT-III | DCT-IV | FWHT |
|---|---|---|---|---|---|---|---|---|
| 16 | 2:1 | 0.9971 | 0.9971 | 0.9876 | 0.9971 | 0.9849 | 0.9733 | 0.9977 |
| | 4:1 | 0.9227 | 0.9225 | 0.9004 | 0.9225 | 0.8756 | 0.8665 | 0.8908 |
| | 8:1 | 0.8149 | 0.8148 | 0.8099 | 0.8148 | 0.7844 | 0.7814 | 0.8102 |
| | 16:1 | 0.7078 | 0.7078 | 0.7103 | 0.7078 | 0.7147 | 0.7149 | 0.7078 |
| 64 | 2:1 | 0.9993 | 0.9992 | 0.9949 | 0.9992 | 0.9932 | 0.9880 | 0.9977 |
| | 4:1 | 0.9254 | 0.9248 | 0.9055 | 0.9248 | 0.8838 | 0.8785 | 0.8908 |
| | 8:1 | 0.8080 | 0.8079 | 0.8074 | 0.8079 | 0.7817 | 0.7784 | 0.8102 |
| | 16:1 | 0.7194 | 0.7194 | 0.7208 | 0.7194 | 0.6921 | 0.6909 | 0.7079 |
| 256 | 2:1 | 0.9995 | 0.9995 | 0.9977 | 0.9995 | 0.9967 | 0.9943 | 0.9977 |
| | 4:1 | 0.8989 | 0.8982 | 0.8859 | 0.8982 | 0.8714 | 0.8685 | 0.8908 |
| | 8:1 | 0.7682 | 0.7682 | 0.7730 | 0.7682 | 0.7521 | 0.7495 | 0.8103 |
| | 16:1 | 0.6707 | 0.6707 | 0.6744 | 0.6707 | 0.6421 | 0.6410 | 0.7079 |
| 512 | 2:1 | 0.9996 | 0.9995 | 0.9984 | 0.9995 | 0.9977 | 0.9960 | 0.9977 |
| | 4:1 | 0.8701 | 0.8696 | 0.8589 | 0.8696 | 0.8498 | 0.8481 | 0.8908 |
| | 8:1 | 0.7245 | 0.7245 | 0.7222 | 0.7245 | 0.7075 | 0.7061 | 0.8103 |
| | 16:1 | 0.6042 | 0.6041 | 0.6032 | 0.6041 | 0.5772 | 0.5764 | 0.7079 |

**Table 6. PESQ (mean value for 40 speech signals)**

| Frame Size | CR | GT-I | GT-II | DCT-I | DCT-II | DCT-III | DCT-IV | FWHT |
|---|---|---|---|---|---|---|---|---|
| 16 | 2:1 | 4.4416 | 4.4310 | 4.1698 | 4.4310 | 4.1185 | 3.5611 | 4.3531 |
| | 4:1 | 3.1942 | 3.1909 | 2.7712 | 3.1909 | 2.3933 | 2.2794 | 2.4149 |
| | 8:1 | 2.0340 | 2.0341 | 1.9917 | 2.0341 | 1.7991 | 1.7417 | 1.8151 |
| | 16:1 | 1.2483 | 1.2483 | 1.3122 | 1.2483 | 1.3430 | 1.3142 | 1.2483 |
| 64 | 2:1 | 4.4977 | 4.4947 | 4.3766 | 4.4947 | 4.3556 | 4.0245 | 4.3528 |
| | 4:1 | 3.6200 | 3.6144 | 3.1373 | 3.6144 | 2.7790 | 2.6943 | 2.4078 |
| | 8:1 | 2.5419 | 2.5414 | 2.4330 | 2.5414 | 2.1844 | 2.1446 | 1.8060 |
| | 16:1 | 1.4852 | 1.4853 | 1.5121 | 1.4853 | 1.5558 | 1.5266 | 1.2382 |
| 256 | 2:1 | 4.4995 | 4.4987 | 4.4417 | 4.4987 | 4.4290 | 4.2345 | 4.3538 |
| | 4:1 | 3.5767 | 3.5683 | 3.2496 | 3.5683 | 2.9468 | 2.8848 | 2.4153 |
| | 8:1 | 2.6815 | 2.6806 | 2.6191 | 2.6806 | 2.3642 | 2.3271 | 1.8157 |
| | 16:1 | 1.8305 | 1.8304 | 1.8811 | 1.8304 | 1.7276 | 1.6998 | 1.2492 |
| 512 | 2:1 | 4.4997 | 4.4994 | 4.4573 | 4.4994 | 4.4478 | 4.2944 | 4.3534 |
| | 4:1 | 3.3688 | 3.3662 | 3.0225 | 3.3662 | 2.7844 | 2.7334 | 2.4121 |
| | 8:1 | 2.5053 | 2.5048 | 2.3404 | 2.5048 | 2.1388 | 2.1041 | 1.8115 |
| | 16:1 | 1.5098 | 1.5097 | 1.4645 | 1.5097 | 1.4017 | 1.3762 | 1.2444 |

## V. CONCLUSION

We proposed two graph structures for audio signals in order to obtain a Graph-base Transform basis matrix. Then we used GT basis matrix for data de-correlation and audio compression. Experimental results show that our method has better performance than other Transform-based methods like DCT and FWHT. Results also show that GT algorithm for music signals has better performance than for speech signals. In this paper we have introduced a static structure for graphs but with a dynamic structure, performance of GT would be better. In future researches we will test dynamic graph structures for better audio compression.

## REFERENCES

[1] K. R. Rao and P. C. Yip, *The transform and data compression handbook*. CRC press, 2000.

[2] J. P. Princen, A. W. Johnson und A. B. Bradley: *Subband/transform coding using filter bank designs based on time domain aliasing cancellation*, IEEE Proc. Intl. Conference on Acoustics, Speech, and Signal Processing (ICASSP), 2161–2164, 1987.

[3] Hou, J., Liu, H., & Chau, L. P. (2016, October). Graph-based transform for data decorrelation. In *Digital Signal Processing (DSP), 2016 IEEE International Conference on* (pp. 177-180). IEEE.

[4] Egilmez, H. E., Said, A., Chao, Y. H., & Ortega, A. (2015, September). Graph-based transforms for inter predicted video coding. In *Image Processing (ICIP), 2015 IEEE International Conference on* (pp. 3992-3996). IEEE.

[5] J. Hou, L. Chau, N. Magnenat-Thalmann, and Y. He, "Human motion capture data tailored transform coding," *IEEE Transactions on Visualization and Computer Graphics*, vol. 21, no. 7, pp. 848–859, 2015.

[6] N. Ahmed, T. Natarajan, and K. R. Rao, "Discrete cosine transform," *IEEE Transactions on Computers*, vol. 100, no. 1, pp. 90–93, 1974.

[7] GUATTERY S., MILLER G. L.: Graph embeddings and Laplacian eigenvalues. SIAM J. Matrix Anal. Appl. 21, 3 (2000), 703–723.

[8] D. Shuman, S. K. Narang, P. Frossard, A. Ortega, P. andergheynst, *et al.*, "The emerging field of signal processing on graphs: Extending high-dimensional data analysis to networks and other irregular domains," *IEEE Signal Processing Magazine*, vol. 30, no. 3, pp. 83–98, 2013.

[9] Rao, K. R., & Yip, P. (2014). *Discrete cosine transform: algorithms, advantages, applications*. Academic press.

[10] X. Shao and S. G. Johnson, "Type-II/III DCT/DST algorithms with reduced number of arithmetic operations," *Signal Processing*, vol. 88, pp. 1553–1564, June 2008.

[11] Martucci, S. A. (1994). Symmetric convolution and the discrete sine and cosine transforms. *IEEE Transactions on Signal Processing*, *42*(5), 1038-1051.

[12] Taal, C. H., Hendriks, R. C., Heusdens, R., & Jensen, J. (2010, March). A short-time objective intelligibility measure for time-frequency weighted noisy speech. In *Acoustics Speech and Signal Processing (ICASSP), 2010 IEEE International Conference on* (pp. 4214-4217). IEEE.

[13] Ksentini, K. P. A., Viho, C., & Bonnin, J. (2009). Perceptual evaluation of speech quality (PESQ): An objective method for end-to-end speech quality assessment of narrowband telephone networks and speech codecs. *University of Rennes*.

[14] Fino, B. J., & Algazi, V. R. (1976). Unified matrix treatment of the fast Walsh-Hadamard transform. *IEEE Transactions on Computers*, (11), 1142-1146.